%% file: ms.tex
\pdfoutput=1
\documentclass[sigconf]{acmart}

\def\BibTeX{{\rm B\kern-.05em{\sc i\kern-.025em b}\kern-.08emT\kern-.1667em\lower.7ex\hbox{E}\kern-.125emX}}

\usepackage{subfig}

\usepackage{xspace}

\newcommand{\schemeA}{ActiveTX-IB\xspace}
\newcommand{\schemeB}{IC-PB\xspace}
\newcommand{\schemeC}{EMB-PB\xspace}
\newcommand{\schemeD}{EMB-IB\xspace}
\newcommand{\sharedBand}{PB\xspace}
\newcommand{\independentBand}{IB\xspace}
\copyrightyear{2019}
\acmYear{2019}
\setcopyright{acmcopyright}
\acmConference[CF '19]{Proceedings of the 16th conference on Computing Frontiers}{April 30-May 2, 2019}{Alghero, Italy}
\acmBooktitle{Proceedings of the 16th conference on Computing Frontiers (CF '19), April 30-May 2, 2019, Alghero, Italy}
\acmPrice{15.00}
\acmDOI{10.1145/3310273.3321555}
\acmISBN{978-1-4503-6685-4/19/05}

\begin{document}
	
	\title{Towards Realistic Battery-DoS Protection of Implantable Medical Devices}
	%\titlenote{Produces the permission block, and
	%  copyright information}
	%\subtitle{Extended Abstract}
	%\subtitlenote{The full version of the author's guide is available as
	%  \texttt{acmart.pdf} document}
	
	\author{Muhammad Ali Siddiqi}
	\orcid{0000-0002-8554-7077}
	\affiliation{%
		\institution{Department of Neuroscience, Erasmus Medical Center}
		\city{Rotterdam}
		\state{The Netherlands}
	}
	\email{m.siddiqi@erasmusmc.nl}
	\author{Christos Strydis}
	\affiliation{%
		\institution{Department of Neuroscience, Erasmus Medical Center}
		\city{Rotterdam}
		\state{The Netherlands}
	}
	\email{c.strydis@erasmusmc.nl}

	\begin{abstract}
		Modern Implantable Medical Devices (IMDs) feature wireless connectivity, which makes them vulnerable to security attacks.
		Particular to IMDs is the battery Denial-of-Service attack whereby attackers aim to fully deplete the battery by occupying the IMD with continuous authentication requests.
		Zero-Power Defense (ZPD) based on energy harvesting is known to be an excellent protection against these attacks.
		This paper establishes essential design specifications for employing ZPD techniques in IMDs, offers a critical review of ZPD techniques found in literature and, subsequently, gives crucial recommendations for developing comprehensive ZPD solutions.
	\end{abstract}
	
	\begin{CCSXML}
		<ccs2012>
		<concept>
		<concept_id>10002978.10003001.10003003</concept_id>
		<concept_desc>Security and privacy~Embedded systems security</concept_desc>
		<concept_significance>500</concept_significance>
		</concept>
		<concept>
		<concept_id>10002978.10003001.10003599</concept_id>
		<concept_desc>Security and privacy~Hardware security implementation</concept_desc>
		<concept_significance>500</concept_significance>
		</concept>
		<concept>
		<concept_id>10002978.10003006.10011610</concept_id>
		<concept_desc>Security and privacy~Denial-of-service attacks</concept_desc>
		<concept_significance>500</concept_significance>
		</concept>
		</ccs2012>
	\end{CCSXML}
	
	\ccsdesc[500]{Security and privacy~Embedded systems security}
	\ccsdesc[500]{Security and privacy~Hardware security implementation}
	\ccsdesc[500]{Security and privacy~Denial-of-service attacks}
	
	%
	% Keywords. The author(s) should pick words that accurately describe the work being
	% presented. Separate the keywords with commas.
	\keywords{Implantable medical device, IMD, energy harvesting, wireless power transfer, zero-power defense, authentication protocol, denial-of-service attack, battery DoS}
	
	\maketitle
	
	\input{paperbody}
	
	\begin{acks}
		This work has been supported by the EU-funded project SDK4ED (Grant Agreement No. 780572) and would not have been complete without the invaluable feedback of Prof.dr.ir. Wouter A. Serdijn and Vasileios Skrekas.
	\end{acks}
	
	\bibliographystyle{ACM-Reference-Format}
	\bibliography{ms}
	
\end{document}

%% file: paperbody.tex
\input{000-introduction}

\input{100-eh_in_imds}

\input{125-using_eh_for_security}

\input{150-design_considerations}

\input{200-zpd_techniques}

\input{500-recommendations}

\input{900-conclusion}

%% file: 000-introduction.tex
\section{Introduction}
\label{sec:introduction}

\begin{figure}[!t]
	\centering
	\includegraphics[trim={4.4cm 5.2cm 4.4cm 4.2cm},clip,scale=0.4]{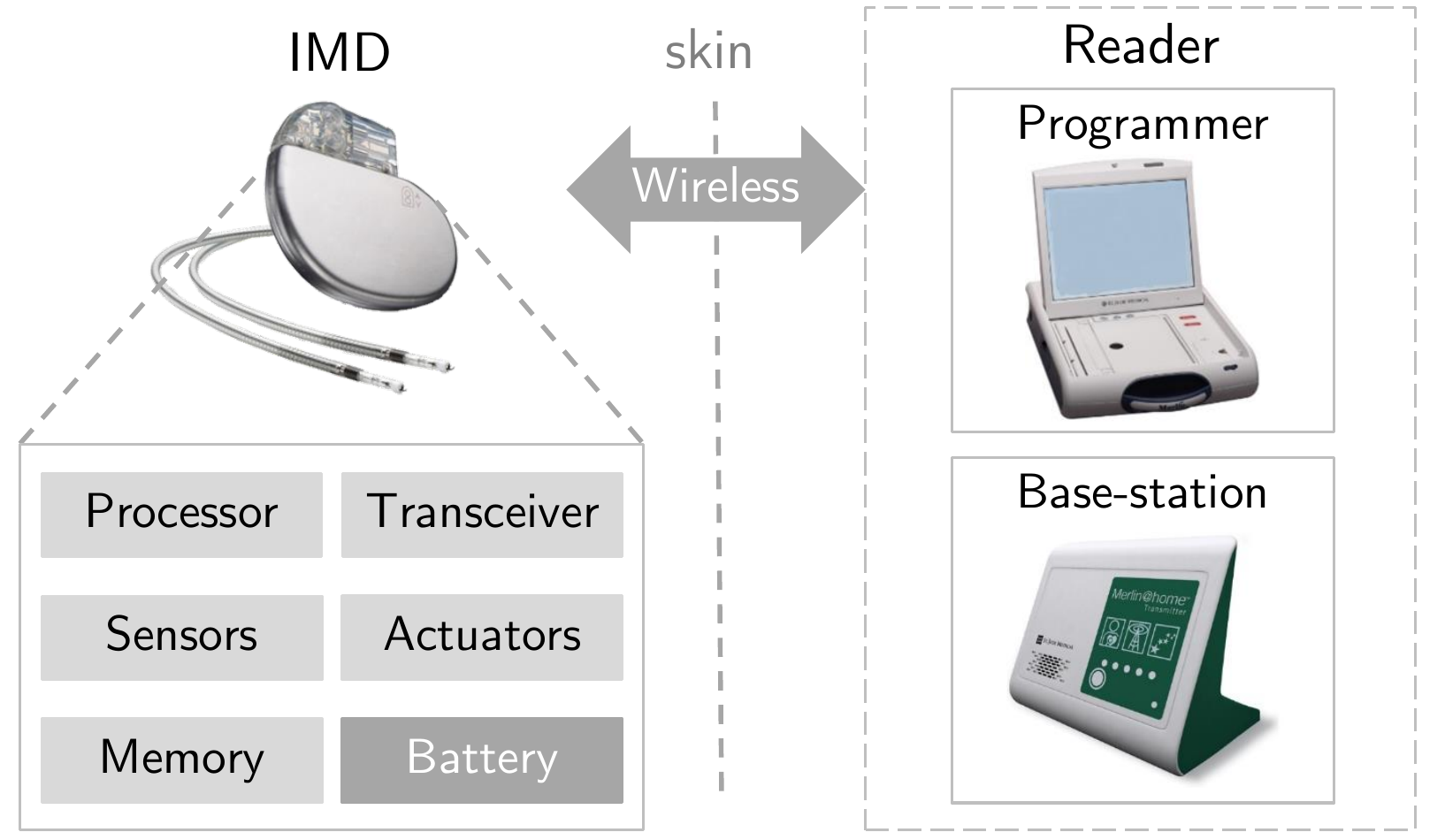}
	\caption{A Reader/IMD system}
	\label{fig:reader-imd-system}
\end{figure}

Implantable medical devices (IMDs) such as cardiac pacemakers, neurostimulators, infusion pumps and more, are autonomous devices with extremely high dependability and safety constraints.
The typical operational lifetime of these battery-powered devices is around a decade or so while implanted in the patient's body.
Almost all of these devices are equipped with wireless connectivity via a transceiver in order to support and complement their treatment capabilities~\cite{siddiqi2018attack}.
They can communicate with an external reader (see Figure~\ref{fig:reader-imd-system}) for e.g. monitoring patient health, updating IMD settings, and so on.
However, despite their benefits, these communication capabilities open the door for malicious entities to wirelessly connect to the device in order to steal private patient data, achieve mis-diagnosis, or even cause physical harm.
An attacker can cause physical harm either by changing the IMD functionality (e.g., by managing to send incorrect commands) or through a Denial-of-Service (DoS) attack.
One such attack is the \emph{battery DoS} where the attacker can force the IMD to continuously run an energy-consuming operation, which ultimately results in power loss and IMD shutdown.
As an example, he/she can repeatedly request the IMD to establish a secure channel using incorrect credentials.
Consequently, the IMD will run part of an energy-consuming authentication protocol for analyzing every request, which will drain the battery.
As indicated in the IMD-threat-modeling analysis in~\cite{siddiqi2018attack}, battery DoS is one of the easiest to mount and highly effective attacks. This is also backed by the majority of the IMD-specific-ethical-hacking efforts in which the batteries of commercial IMDs were depleted using black-box approaches~\cite{halperin2008pacemakers,marin2016security}.

The only robust way of protecting an IMD against a battery DoS is by running the above-mentioned (energy-consuming) authentication operation using only \emph{free} harvested energy.
It can be argued that there is no necessity for this zero-power defense (ZPD) mechanism since technology exists to wirelessly charge IMD batteries when they are running low (as discussed in Section~\ref{sec:eh_in_imds}). However, this recharging feature is only available in less critical IMDs, such as spinal-cord stimulators.
For critical devices such as pacemakers, there is a reluctance among the medical community to give recharging responsibility to the patients, in order to avoid patient errors. Moreover, the physicians prefer to replace the whole IMD after a certain period to get the latest technology~\cite{lafrance2014who}.
Besides, even by assuming that all IMDs have this capability, the attacker can still drain the battery before the patient or doctor has a chance to recharge it.

Energy harvesting is a widely used concept employed in a variety of devices including RFIDs.
However, ZPD for IMDs introduces new challenges that do not apply in other domains. 
This paper is the first to facilitate the transition from \emph{concept} to \emph{practical} ZPD designs for IMDs. Based on a clear-cut set of design considerations, we survey and evaluate the current state of the art and proceed to propose specific recommendations for enhancing existing IMDs.
Essentially, this work makes the following novel contributions:
\begin{itemize}
	\item We consolidate ZPD design considerations for the specific domain of IMDs.
	\item We perform a survey of existing systems and highlight their limitations based on the above considerations.
	\item We provide recommendations in order to develop comprehensive protection of IMDs against battery-DoS attacks.
\end{itemize}

The rest of the paper is organized as follows. We provide brief background on the use of energy harvesting in IMDs in Section~\ref{sec:eh_in_imds}, and then provide motivation for using it to enhance IMD security in Section~\ref{sec:using_eh_for_security}.
In Section~\ref{sec:design_considerations}, we provide detailed ZPD design considerations.
Based on these considerations, we review and evaluate state-of-the-art ZPD solutions in Section~\ref{sec:zpd_techniques}.
In Section~\ref{sec:recommendations}, we provide recommendations for improving ZPD designs.
We conclude the discussion in Section~\ref{sec:conclusion}.

%% file: 100-eh_in_imds.tex
\section{Energy Harvesting in IMDs}
\label{sec:eh_in_imds}

The use of energy harvesting in IMDs is not new. 
The application of this concept, however, has been very narrow in this domain, i.e., in wireless power transfer (WPT)\footnote{The term \emph{energy harvesting} generally refers to harvesting energy from ambient sources, whereas \emph{WPT} refers to the intentional transfer of energy from a dedicated charging device~\cite{costanzo2014electromagnetic}. In this paper, we use the terms interchangeably.} to recharge IMD batteries.
For instance, there are several rechargeable neurostimulators that are commercially available~\cite{medtronic2018restore,abbott2018prodigy}.
In this specific category of implants, there is a rising trend towards increased IMD-power requirements due to recent advances in neuromodulation-related pain relief.
For such power-hungry devices, a non-rechargeable battery would result in a very short IMD lifespan and subsequently require expensive surgeries in order to replace the battery-depleted implants. One way of avoiding this is to use larger battery sizes, which can quickly become impractical to implant.
Hence, the natural solution is to use rechargeable systems, which can prevent the need for frequent surgeries and would result in smaller battery sizes and implants as a whole~\cite{mehta2018when}.

%% file: 125-using_eh_for_security.tex
\section{Energy Harvesting for IMD security}
\label{sec:using_eh_for_security}

As indicated in Section~\ref{sec:introduction}, battery DoS is one of the easiest to mount and highly effective attacks.
In light of the fact that energy harvesting has already been employed by some classes of IMDs, the use of this concept, in the form of ZPD, has now become quintessential to protecting all IMDs against battery DoS. 
In this scheme, the IMD, while authenticating the external entity that is trying to communicate, can run the energy-consuming security primitives using the RF energy harvested from the incoming communication messages. 
The IMD is allowed to use the battery for subsequent operations \emph{only} after the entity is authenticated.
This prevents the IMD depleting its battery to entertain continuous bogus messages from a malicious entity.

%% file: 150-design_considerations.tex
\section{Design Considerations}
\label{sec:design_considerations}

In this section, we enumerate and discuss various considerations that should be taken into account when approaching the design of an IMD-specific ZPD system.

\subsection{Choice of WPT technique}
\label{sec:wpt_techniques}

Since ZPD is based on the concept of wireless energy harvesting, it is important to briefly discuss the WPT techniques that enable such strategies.
A typical WPT setup is shown in Figure~\ref{fig:generic_wpt}~\cite{kadirvel2012power, mansano2016autonomous}.
State-of-the-art IMD-specific WPT techniques can be broadly categorized into three types\footnote{Note that this classification is not universal.}~\cite{basaeri2016review}:

\begin{figure}[!t]
	\centering
	\includegraphics[trim={0.5cm 0.5cm 0.5cm 0.5cm},clip,scale=0.51]{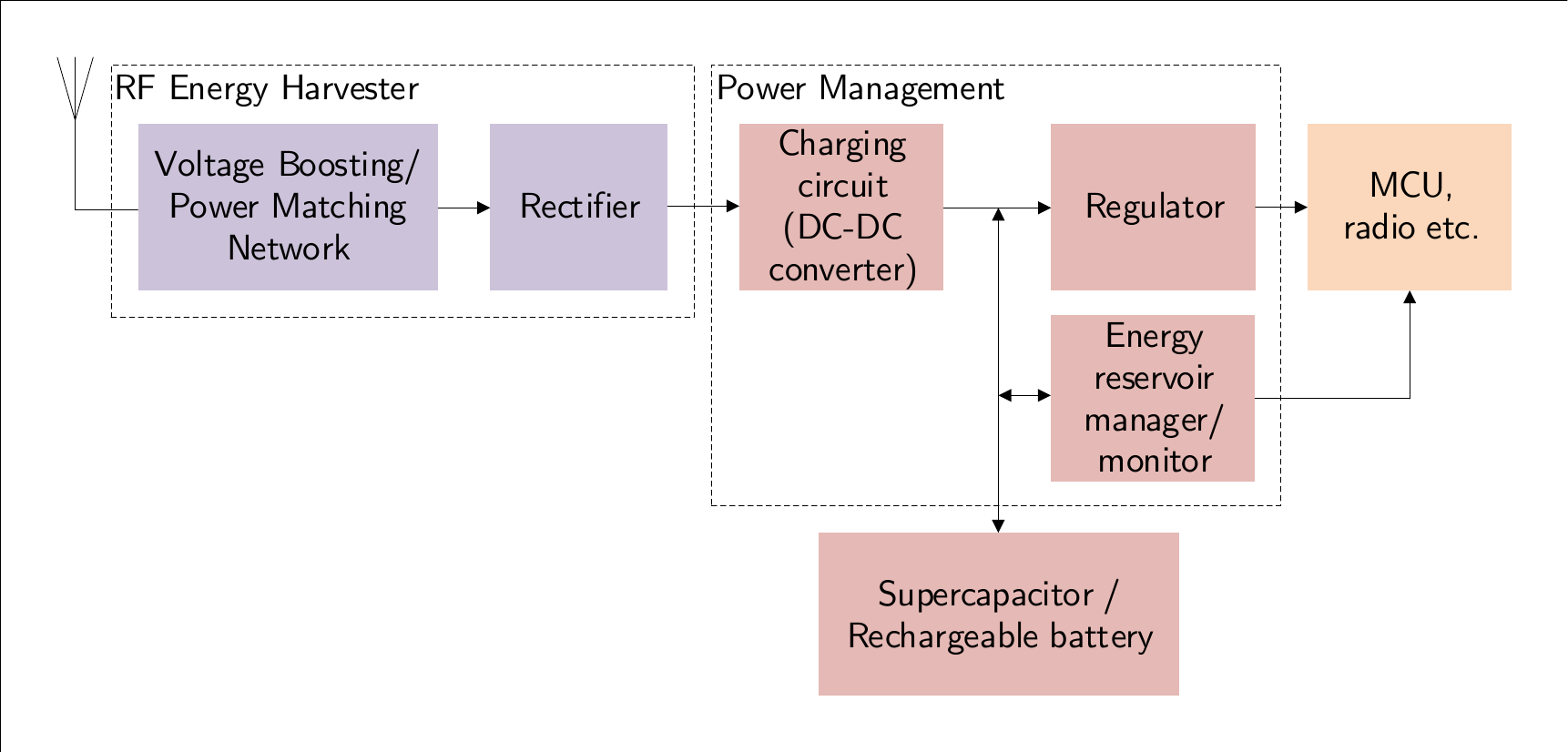}
	\caption{A typical WPT System (RF Energy Harvesting)}
	\label{fig:generic_wpt}
\end{figure}

\begin{table}
\caption{Comparison of WPT techniques}
\label{table:wpt_summary}
\centering
\footnotesize
\begin{tabular}{lcccc}
	\hline
	Technique 		& Range & Biological& Transferred& Receiver \\
	&  		& effects			& power 	& size	\\
	\hline
	IPT 			&   $-$&   $-$	&   $+$ & $-$ \\
	RFPT	 		&   $+$	&   $-$	&	$-$	& $+$ \\
	APT		 		&   $-^*$&  $+$	&	$+$	& $+$ \\
	\hline
	\multicolumn{5}{l}{$+/-$: relatively good/poor performance, $^*$: requires (non-air) medium} \\
\end{tabular}
\end{table}

\subsubsection{Inductive Coupling (IC)}

Near-field WPT is usually categorized as inductive coupling or inductive power transfer (IPT). IPT usually involves the use of two coupled coils that have the same inductance. 
The transmitter coil is placed outside the body. When an AC current passes through it, voltage is induced due to electromagnetic induction in the receiver coil, which is located inside the body.
IPT is the dominant method that is used to wirelessly recharge commercial IMDs, specifically neurostimulators~\cite{abbott2018prodigy,medtronic2018restore}.

\subsubsection{Radio Frequency (RF)}

If the transfer is in the transition region (mid field)~\cite{ho2014wireless} or far field then the WPT system is usually categorized as RF power transfer (RFPT).
Here, antennas are not just limited to coils for the transmission of power.
A typical RFPT system is shown in Figure~\ref{fig:generic_wpt}.

\subsubsection{Acoustic/Ultrasound}

This WPT category harvests acoustic waves, which are usually from ultrasound frequencies. In acoustic power transfer (APT), the transmitter node, while in contact with the skin, generates these waves using a piezoelectric transducer. These waves induce voltage on a piezoelectric device in the receiver node, which is located inside the body along with the IMD.

The advantages and drawbacks of the three WPT techniques are summarized in Table~\ref{table:wpt_summary} in terms of operating range, potential biological effects, amount of transferred power and receiver area.
The choice of WPT scheme and associated transferred-power amount has an impact on the real-time IMD \emph{performance}, and also on the \emph{size} of the energy reservoir and, subsequently, the IMD as a whole.
This is further discussed in the subsequent sections.

\subsection{Medical safety constraints}

The ZPD technique should satisfy the various requirements by the FDA, FCC, IEEE etc., in order to prevent any adverse biological effects on human tissue due to excess electromagnetic-energy exposure.
IEEE puts constraints on the intensity of RF signals and defines maximum-permissible-exposure (MPE) limits for magnetic and electric fields~\cite{ieee2006standard}.
In addition to RF-signal intensity, the signal frequency has a significant impact on the amount of energy absorbed in the human tissue and the resulting potential to cause harm. This absorption is characterized by \emph{specific absorption rate} (SAR), which is expressed in $\frac{W}{kg}$ or $\frac{mW}{kg}$.
The peak-spatial-average SAR values for exposure of the public and controlled environments are 2 $\frac{W}{kg}$ and 10 $\frac{W}{kg}$, respectively (over 10 $g$ of tissue)~\cite{ieee2006standard}.
FDA also has guidelines regarding intensity of acoustic signals in $\frac{W}{cm^2}$, namely \emph{spatial peak temporal average intensity} ($I_{SPTA}$) and \emph{spatial peak pulse average intensity} ($I_{SPPA}$)~\cite{fda2008guidance}.
Satisfying these constraints impacts the choice of WPT scheme (as discussed in Section~\ref{sec:wpt_techniques}).

\subsection{Frequency-band constraints}
\label{sec:freq_band_constraints}

Certain FCC constraints also need to be met in order to avoid interference with the devices operating in the same frequency band. For example, the MedRadio band, which is reserved for IMD communication, does not allow an equivalent isotropically radiated power (EIRP) of more than 25 $\mu W$~\cite{fcc2018medradio}. Since this amount of power is unrealistic for WPT, a separate band should be used for power transfer, whereas the MedRadio band can be used for data communication.
This implies increased cost and size due to the use of two antennas. One solution could be to use a single ISM-band (13.56 $MHz$) antenna for both WPT and data communication, however this would result in lower data rates due to smaller allowed bandwidth than that of MedRadio~\cite{martins2017energy}.

\subsection{Harvested vs. Consumed power}
\label{sec:harvested_vs_consumed}

Harvested energy needs to stay above the consumed power in order for the energy consumers to work seamlessly. 
Otherwise, an energy reservoir should be employed so that it can collect sufficient energy before the IMD can use it.
Technically, due to this reservoir, the ZPD scheme should always work, but the charging delay limits usability and real-time behavior, which can be critical in the case of emergencies.

\subsection{Choice of energy reservoir}
Either a supercapacitor (supercap) or a rechargeable battery can be employed as the energy reservoir.
Supercaps in general have a longer lifespan and support more recharge cycles than batteries~\cite{guerra2016}, and thus are more suitable for IMDs.
The supercap, if used, could limit the range of applied charging voltage, since these components have low operating-voltage limits.
Also, as indicated by~\cite{silabs2013}, the capacitor size has to incorporate the losses due to the decoupling capacitors connected to the energy consumers.

\subsection{Passive wireless communication}
\label{sec:taxonomy}

\begin{figure}[!t]
	\centering
	\includegraphics[trim={0.5cm 0.5cm 0.5cm 0.5cm},clip,scale=0.55]{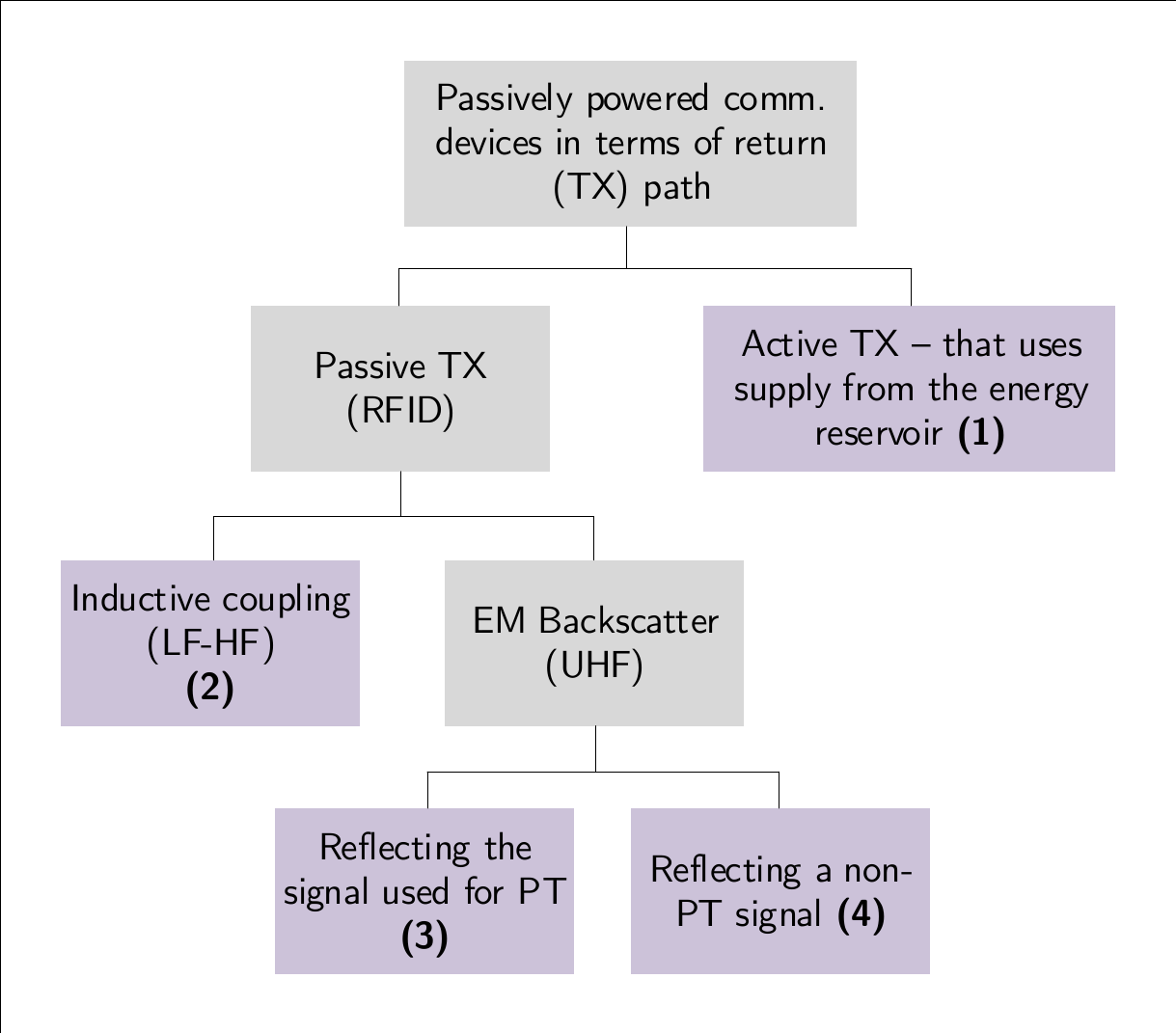}
	\caption{Classification of passive communication devices in terms of transmitter implementation}
	\label{fig:zpd-taxonomy}
\end{figure}

Passive communication relies on WPT schemes in order to function without the need of an on-board battery. This concept forms the basis of ZPD strategies, which will be discussed in Section~\ref{sec:zpd_techniques}. The most critical component of these passive devices is the wireless transceiver that can cause significant peak power consumption based on the design choice. Based on the choice of transmitter, which subsequently impacts the receiver implementation, we categorize these devices into four schemes, as depicted in Figure~\ref{fig:zpd-taxonomy}. The different schemes at the leaf nodes are numbered accordingly and are subsequently explained.
The first part of the scheme name indicates the type of wireless communication whereas the suffix indicates whether the communication shares the power-transfer-signal frequency band (\sharedBand) or uses an independent band (\independentBand).

\begin{figure}
	\centering
	\subfloat[\schemeA]{\label{fig:wpt-scheme1}{\includegraphics[trim={0.5cm 0.5cm 0.5cm 0.5cm},clip,scale=0.7]{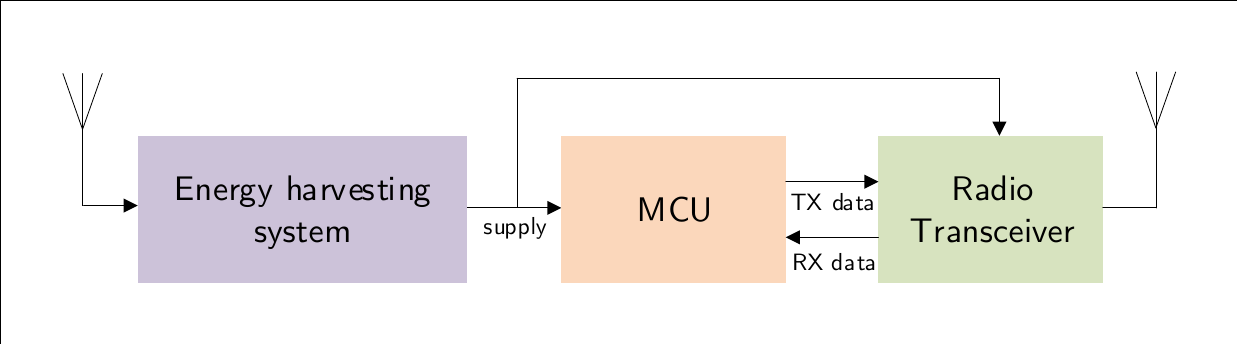} }}%
	\qquad
	\subfloat[\schemeB and \schemeC]{\label{fig:wpt-scheme2}{\includegraphics[trim={0.5cm 0.5cm 0.5cm 0.5cm},clip,scale=0.7]{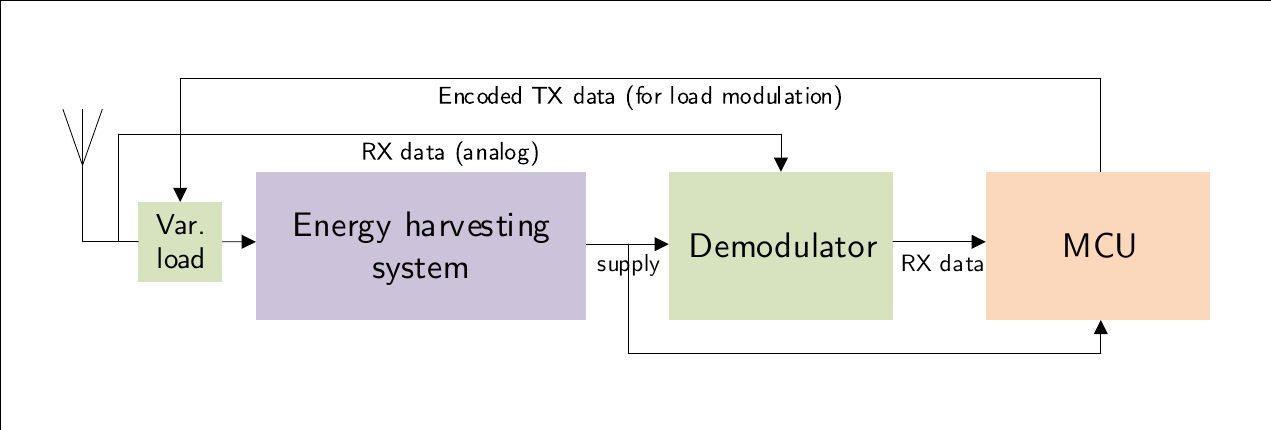} }}%
	\qquad
	\subfloat[\schemeD]{\label{fig:wpt-scheme4}{\includegraphics[trim={0.5cm 0.5cm 0.5cm 0.8cm},clip,scale=0.7]{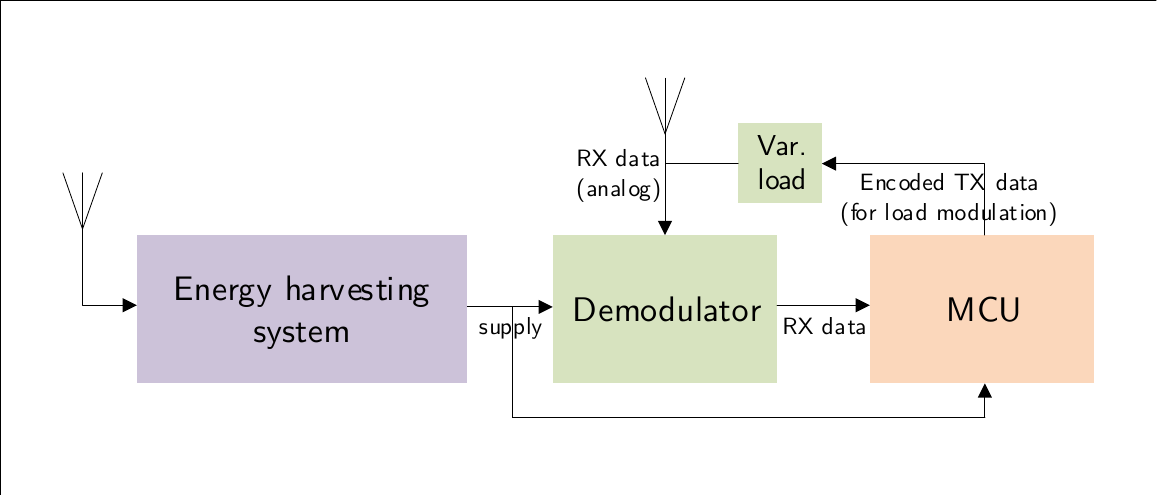} }}%
	\caption{Schematics of different passive communication schemes for ZPD}%
	\label{fig:wpt-schemes}%
\end{figure}

\subsubsection{\schemeA}

The passive device has an \emph{active} transceiver, i.e., it actively transmits (using supply from the energy reservoir) instead of reflecting the incident RF signal, as shown in Figure~\ref{fig:wpt-scheme1}. This scheme is employed by the design in~\cite{schaumont2016secure}.

\subsubsection{\schemeB}

The forward direction (reader to passive device) communication uses the same signal that is used for inductive power transfer, which lies in the low- or high-frequency band (LF-HF). For the reverse direction, the electrical properties of the inductive coil are changed (by load modulation, in this case \emph{Load Shift Keying}), which affects the same inductive coupling field, and is thus detected by the reader (see Figure~\ref{fig:wpt-scheme2}). The design in~\cite{chang2017power} employs this scheme.

\subsubsection{\schemeC}

Compared to the previous scheme, RF/Electro\-magnetic backscattering (EMB), which reflects the incident RF, is used for data transmission instead of inductive coupling. Here, the incident RF is used for both energy harvesting and data communication (see Figure~\ref{fig:wpt-scheme2}). The RF is reflected if the load across the antenna feed-point is minimum, and vice versa. One of the works that employ this scheme is~\cite{sample2007design}.
The use of EMB helps eliminate the high peak power consumption of a conventional RF transmitter. 
This is important for passive devices because even to transmit just a few bits of data, the peak power may exceed the incoming power, which will result in device malfunction in the absence of a reservoir. Note that the use of EMB for transmission is fully beneficial only if a simple and low-power circuit is used for the receive path, such as an Amplitude-Shift-Keying (ASK) envelop detection.

\subsubsection{\schemeD}

Compared to \schemeC, here the difference is that the WPT signal is different than the one used for EMB (as shown in Figure~\ref{fig:wpt-scheme4}). The design in ~\cite{mansano2016autonomous} uses this scheme.

\schemeA and \schemeD offer the most flexibility since they use separate antennas for WPT and data communication.
As discussed in Section~\ref{sec:freq_band_constraints}, these configurations are helpful in meeting the FCC constraints while maintaining both the sufficient power transfer and data rates.
On the other hand, \schemeB and \schemeC are more economical in terms of resources since they only employ one antenna~\cite{martins2017energy}. This is, however, at the cost of reduced flexibility in terms of data rate.

\subsection{Fundamental security services}
ZPD schemes primarily address \emph{Availability} from the CIANA security services~\cite{siddiqi2018attack}: \emph{Confidentiality}, \emph{Integrity}, \emph{Authentication}, \emph{Non-repudiation} and \emph{Availability}.
Ensuring the first four services can have an indirect impact on Availability.
As an example, if the IMD has a dedicated processor that is responsible for authenticating an external entity, the peak-power consumption of the implant will increase when this peripheral is active. As a result, the bogus messages sent by an attacker will draw more energy from the battery than in the case of a less-secure IMD.
Hence, ensuring one service should not be at the expense of the other.

The choice of cryptographic primitives, which are needed to provide these services, plays a critical role in the design of the energy-harvesting circuit. 
For example, lightweight block ciphers are preferred candidates for achieving data confidentiality because of their low energy profile.
Moreover, in order to achieve integrity and authentication, a cipher-based Message Authentication Code (MAC) should be used instead of a hash-based MAC (HMAC) because of lower energy consumption in software implementations. For dedicated hardware implementations, however, this does not always hold~\cite{patrick2016role}.
Furthermore, for these systems \emph{mutual} authentication should be employed instead of just authenticating the reader unilaterally. This is required to prevent spoofing attacks on the reader~\cite{strydis2013system}.
This implies that the harvested energy should be able to support \emph{both} transmission and reception of data.

\subsection{Emergency access}
In the case of emergencies, the paramedics or first responders should have seamless and fast access to the IMD, without compromising patient safety and security. Hence, an appropriate balance should be attained between usability, safety and security.
It is of paramount importance that the choice of WPT and the associated energy reservoir results in acceptable charging delay in order to ensure real-time performance. Otherwise, it will block legitimate access to the IMD in emergency scenarios.

\subsection{Design suitability}

Existing IMD designs take a long time from concept to market due to pedantic regulatory hurdles.
Therefore, any new ZPD solution should fit in seamlessly in the existing designs resulting in minimal changes and short review cycles.
For example, as mentioned in Section~\ref{sec:harvested_vs_consumed}, technically-speaking a large energy reservoir should always work but this increases the size of the ZPD solution and introduces unnecessary delay, which impacts suitability.

\subsection{Conformity to touch-to-access principle}
\label{sec:touch_to_access}

Any ZPD scheme shall ensure that only the entity in close proximity to the patient for a prolonged period of time is allowed to access the IMD.
This \emph{touch-to-access} principle assumes that it is infeasible for the attacker to get in close proximity since the patient would reject physical contact with untrusted entities~\cite{rostami2013heart,siddiqi2018attack}.

\subsection{Range of operation}

The ZPD solution shall be able to work correctly independently of the implantation depth.
Appropriate balance should be attained between the WPT and the associated thermal effects and energy absorption in the human tissue.
Also, the ZPD solution shall allow the provision of a bedside-base-station operation for the convenience of the patient (see Figure~\ref{fig:reader-imd-system}). This device by definition can be less than 10 feet away from the patient~\cite{merlin2015faq}. However, in order to conform to the touch-to-access principle, this communication should be strictly limited to the bedside range (less than 5 feet away).

%% file: 200-zpd_techniques.tex
\section{A Survey of Existing ZPD Techniques}
\label{sec:zpd_techniques}

\begin{table*}
	\caption{Summary of ZPD strategies}
	\label{table:zpd_summary}
	\centering
	\footnotesize
	\begin{tabular}{lllllll}
		\hline
		Technique 				& Halperin et al.~\cite{halperin2008pacemakers} & Liu et al.~\cite{liu2010secure} & Strydis et al.~\cite{strydis2013system} & Ellouze et al.~\cite{ellouze2013securing,ellouze2018powerless} & Yang et al.~\cite{yang2014chip} & Chang et al.~\cite{chang2017power} \\
		\hline
		%		\midrule
		Satisfy safety constraints	&   -						& -  		&	-					& -							& -			& -	\\
		Satisfy freq. band constraints & -						& Yes 		&	-					& Yes						& Yes		& -	\\
		Harvested vs. consumed power& -							& -			&   -					& -  						& -			& - \\
		Real-time performance		& Yes						& -  		&   -					& -  						& -			& Yes  \\
		Energy reservoir			& Not used					& Not used	&	-					& Not used					& Not used	& - \\
		Type of WPT technique 		& RFPT						& RFPT (ISM)&	-					& RFPT						& IPT		& IPT\\
		Passive wireless communication:&						&   		&						&							&			&\\
		\quad \textit{Scheme}		& \schemeC					& -  		&	-					& \schemeC					& \schemeB	& \schemeB\\
		\quad \textit{ZPD Receive path}& ASK					& -  		&	-					& ASK						& IC (ASK)	& IC (FM)\\
		\quad \textit{ZPD Transmit path}& EMB					& No		& 	-					& EMB 						& IC (ASK)	& IC (ASK)\\
		Security services related:	&   						&   		&						&							&			&	\\
		\quad \textit{Employed primitives}& DE (RC5)			& -  		&	DE, CMAC (MISTY1)	& DE, HMAC					& Hash function	& AES-CCM-128 \\
		\quad \textit{Mutual authentication}& No  				& No  		&	Yes					& Yes						& Yes		& -	\\
		\quad \textit{Avoids pre-shared keys}& Yes (using audio channel)& No& 	No					& Yes						& No		& -	\\
		\quad \textit{Documented vulnerabilities}&\cite{halevi2010pairing}&-&	-					& \cite{rostami2013balancing}& -		& -	\\
		Emergency access	 		& Yes						& No  		& 	No					& Yes						& No		& -	\\
		Touch-to-access		 		& Yes  						& -  		&	No					& Yes						& -			& Yes \\
		Operating distance:			&   						&   		&						&							&			&	\\
		\quad \textit{Max. implantation depth}& 1 cm (animal tissue)& -		&	-					& -							& -			& 2.5 cm (air) \\
		\quad \textit{Bedside-base-station operation}& No				& No		&	No					& No						& No		& No \\
		Design suitability (size,	& WISP and					&RFID module&Security co-processor	& WISP						& -			& -	\\
		additional hardware etc.)	& piezoelectric transducer	& 			& 						&   						& 			&   \\
		\hline
		\multicolumn{7}{l}{`-': Lacking information, DE: Data Encryption, Not used: Advantageous avoidance of an additional component, Yes/No: Satisfies requirement (or not)} \\
		%		\bottomrule
	\end{tabular}
\end{table*} 

In light of the design considerations mentioned in Section~\ref{sec:design_considerations}, we now survey works from literature and discuss their limitations.
We hope that this survey will help us construct more complete solutions.
These works are presented in chronological order, which, to the best of our knowledge, are the only works pertaining to ZPD for IMDs.

Halperin et al.~\cite{halperin2008pacemakers} presented the pioneering work of RFID-style energy harvesting for zero-power defense of IMDs.
They use an RFID module called WISP~\cite{sample2007design}, which employs EMB for the data transmission from the implant to the reader, and simple ASK-envelop detection in the reverse direction, while using RFPT for wireless power transfer.
Their scheme, however, does not perform mutual authentication and its acoustic-communication-based key transport is susceptible to attack, as shown in~\cite{halevi2010pairing}.

The scheme from Liu et al.~\cite{liu2010secure} is the only ZPD work that takes FCC regulations into consideration. They employ the ISM band for RFPT and the MedRadio band for data communication.
It employs a dedicated passive RFID wake-up module, which performs RF-energy harvesting from the incoming signal in order to authenticate the other entity.
Upon successful authentication, the main module is woken up.
This scheme uses pre-shared keys between the reader and the IMD, which makes emergency access impossible.
This is because in emergencies, the IMD and the paramedic reader are likely unknown to each other and therefore do not share a key.

Strydis et al.~\cite{strydis2013system} propose an IMD architecture that isolates the implant functionality from the security tasks by using dedicated processing cores for the respective applications.
They designed the security co-processor from scratch, which was optimized for executing the MISTY1 cipher in terms of energy and performance.
The choice of this dual-core architecture helps in dealing with repeated communication requests that may prevent the implant from performing its primary task. Battery DoS is tackled by ensuring that the security core and the transceiver run on harvested RF energy before mutual authentication of reader/IMD. After successful authentication, these modules are allowed to use battery power for subsequent communication.
However, they did not present a full system implementation.

Ellouze et al.~\cite{ellouze2013securing, ellouze2018powerless} propose an RFID-based, energy-harvesting solution, that uses the same WISP module as employed by~\cite{halperin2008pacemakers}.
In contrast to~\cite{halperin2008pacemakers}, their solution additionally provides mutual authentication.
They use cardiac-signal-based biometrics for authentication and the generation of session keys.
However, the fuzzy-vault-inspired protocol (OPFKA)~\cite{hu2013opfka} employed in their scheme is vulnerable to attacks as demonstrated in~\cite{rostami2013balancing}.

Yang et al.~\cite{yang2014chip} use IPT, and employ the same coil for power transfer and data communication.
Their scheme provides mutual authentication. However, it employs pre-shared keys, and is thus unable to support emergency access.
Moreover, they did not implement a unified ZPD-system since the hash-based authentication was verified separately on an FPGA.

Chang et al.~\cite{chang2017power} propose a generic ZPD solution that is not specific to IMDs per se, however, it covers a spectrum of devices that have more or less the same profile. They propose IPT for the power transfer from the reader. 
This signal is also used for bi-directional communication.
However, they do not give any description of the employed security protocol.

Table~\ref{table:zpd_summary} compares the above ZPD techniques based on the various parameters and design considerations highlighted in Section~\ref{sec:design_considerations}. 
We can see that all listed works lack the evaluation of hazardous biological effects of the employed WPT schemes.
They also do not consider the possibility of a bedside-base-station operation, which is a rising trend in the reader/IMD systems.
Moreover, all the techniques offer insufficient security-services and/or have security vulnerabilities in one form or another.

%% file: 500-recommendations.tex
\section{Recommendations}
\label{sec:recommendations}

We, next, provide recommendations on how existing solutions can be improved in order to better meet the design constraints highlighted in Section~\ref{sec:design_considerations}.

\subsection{Adaptive ZPD}
In modern IMD setups, in addition to the doctor's programmer, we also have a bedside base-station, as shown in Figure~\ref{fig:reader-imd-system}.
For the convenience of the patients, these wireless devices are required to communicate with the IMD from a few feet away~\cite{merlin2015faq}.
With this constraint, IPT- and APT-based ZPD cannot be used for the base-station/IMD authentication.
Hence, with this setup, it is advantageous to employ RFPT for energy harvesting, since they are more flexible compared to IPT and APT in terms of range.
Though the amount of power transferred through RFPT is significantly small compared to IPT/APT, it is not an issue in this specific case since the base-station communication is only used for non-critical daily monitoring.
As a result, this setup can afford long delays due to energy-reservoir charging.
In light of the above, an adaptive ZPD approach should be considered, that e.g., uses IPT/APT for doctor-programmer/IMD communication, and switches to RFPT for base-station/IMD communication.
In terms of implementation cost, it is more economical to use IPT for programmer/IMD communication instead of APT.
This is because same coils can potentially be employed for near-field (programmer communication) and far-field (base-station communication).
On the other hand, the use of APT (for programmer communication) would require the use of piezoelectric transducers in addition to the RF antenna (needed for base-station communication).

\begin{figure}[!t]
	\centering
	\includegraphics[trim={0.6cm 3.4cm 0.9cm 0.2cm},clip,scale=.38]{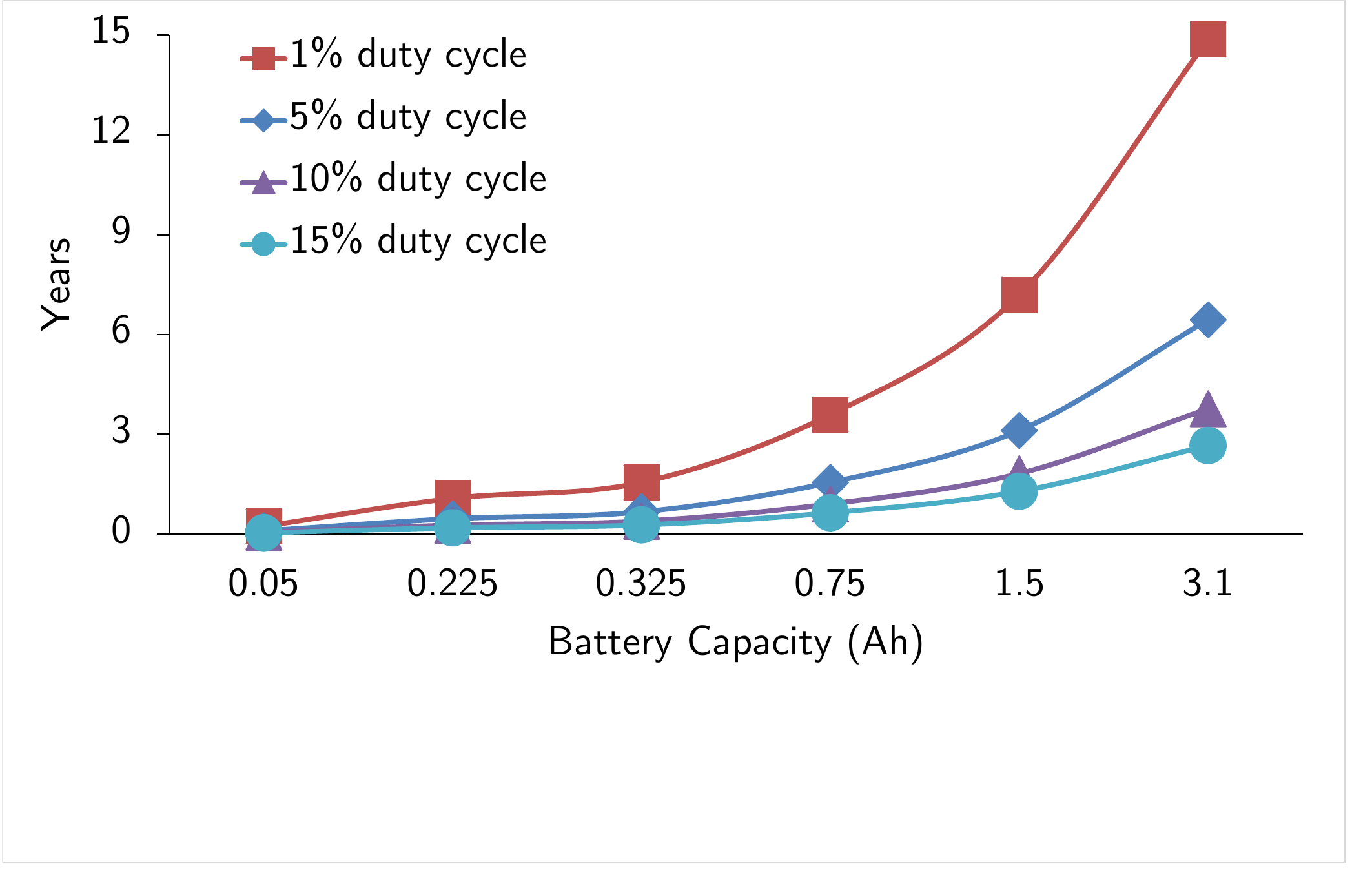}
	\vspace{-0.2cm}
	\caption{IMD-battery lifetime with respect to example processor duty cycles while the transceiver is active for 3 minutes per 24 hours}
	\label{fig:battery-lifetime}
	\vspace{-0.2cm}
\end{figure}

\begin{figure}[!t]
	\centering
	\includegraphics[trim={0.6cm 3.4cm 0.9cm 0.2cm},clip,scale=.38]{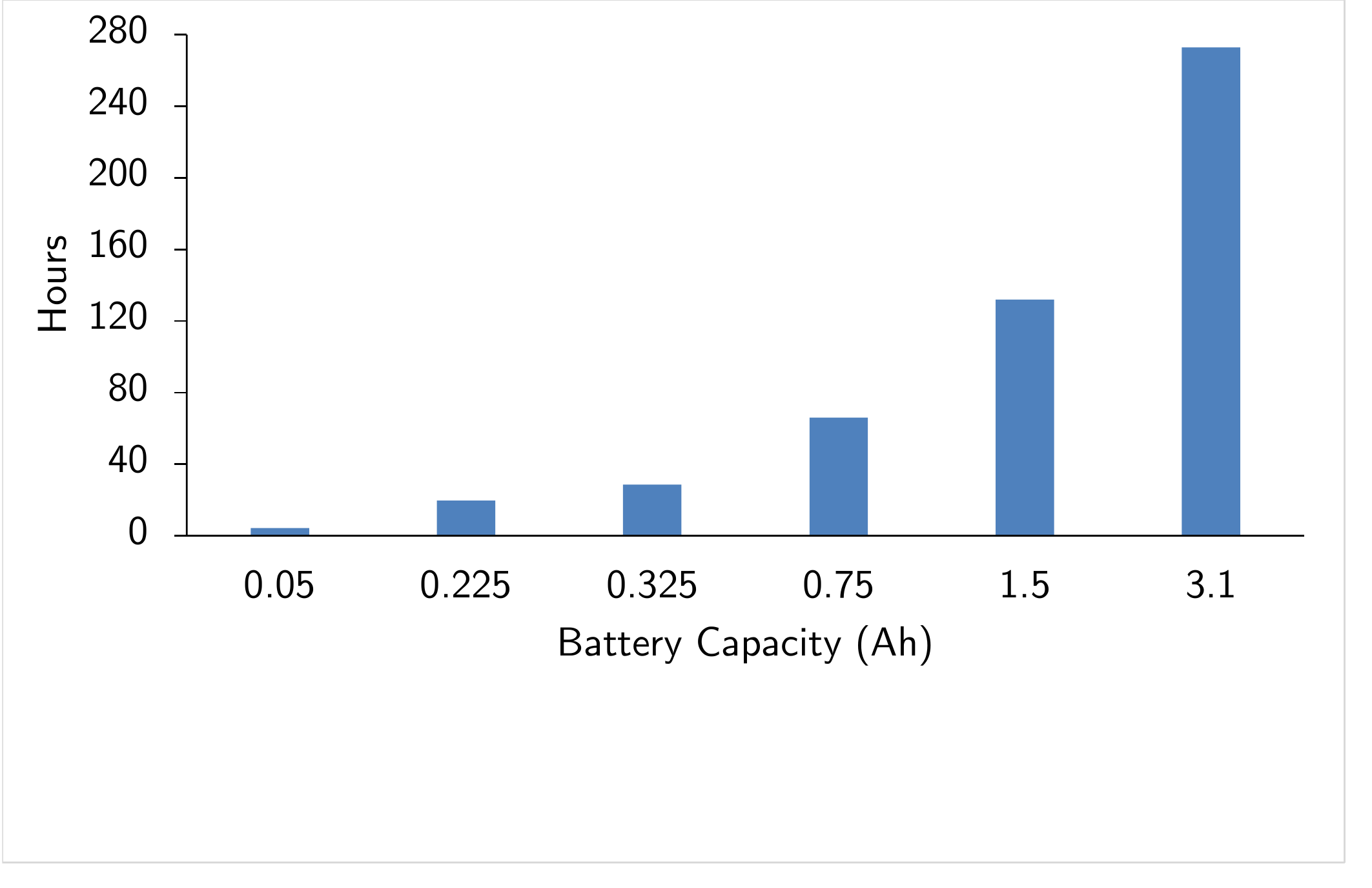}
	\vspace{-0.2cm}
	\caption{Time required to completely deplete a half-full IMD battery through battery DoS}
	\label{fig:battery-dos-attempts}
\end{figure}

\subsection{Main-implant-battery size}
\label{sec:battery_size}

We now discuss how realistic it is to achieve battery DoS when considering actual IMD battery sizes.
The IMD-battery-lifetime trends with respect to example processor duty cycles are shown in Figure~\ref{fig:battery-lifetime}.
For instance, the pacemaker design in~\cite{lindqvist2005compression} has a processor duty cycle of 5\%.
For the calculations, it is assumed that the IMD has a state-of-the-art ultra-low-power ARM Cortex-M0+ based 32-bit MCU~\cite{tinygecko}, running at 19 $MHz$, and an implantable-grade radio transceiver~\cite{microsemi70103}, with an effective data rate of 265 $kbps$.
The duty cycle of the transceiver is assumed to be 0.21\%, which corresponds to 3 minutes of active data communication per 24 hours with a bedside base-station~\cite{merlin2015faq}. The data points correspond to actual implantable-grade battery sizes~\cite{eaglepicher}.
The time required to completely deplete the IMD battery through battery DoS is illustrated in Figure~\ref{fig:battery-dos-attempts}.
On average, we assume half the charge available in the batteries due to normal use.
We also assume that the authentication steps are executed on active modes of the MCU and the transceiver with the current consumption of 0.78 $mA$ and 4.9 $mA$, respectively.
It can be deduced from these plots that, as a first layer of defense, the battery sizes for critical applications, such as pacemakers, should be as large as possible.

\begin{figure}
	\centering
	\includegraphics[trim={0.5cm 0.6cm 0.5cm 0.6cm},clip,scale=0.69]{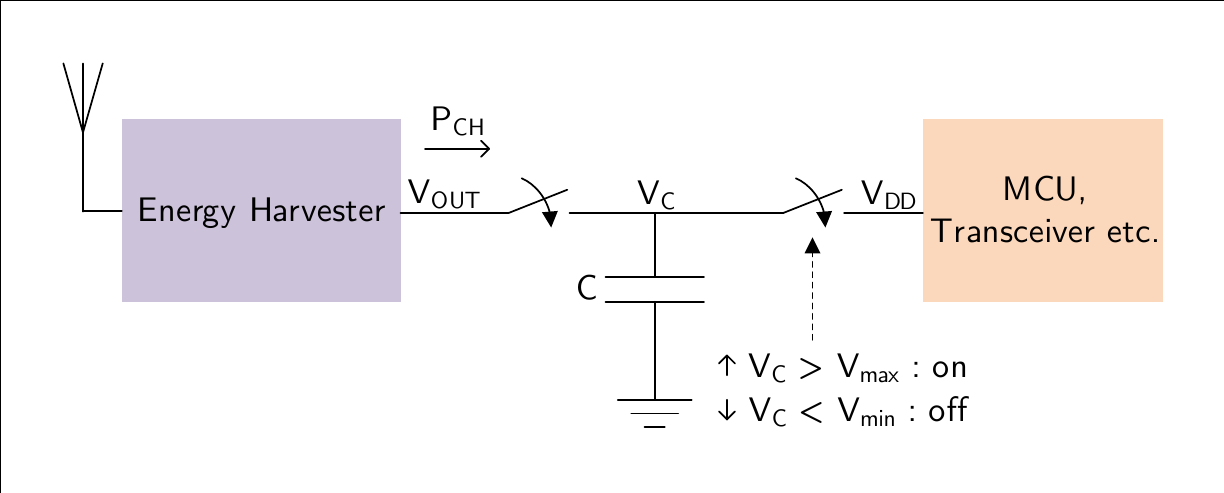}
	\vspace{-0.2cm}
	\caption{Simple ZPD configuration}
	\label{fig:imd-zpd-simple}
\end{figure}

\subsection{Reservoir size and charging delay}
\label{sec:charging_time}

If the peak power of the load is always less than the harvested power then we do not need a reservoir. Otherwise, the size of the reservoir is determined by looking at the required energy consumption of all the consumers during the authentication operation.
Moreover, if a reservoir is required, then it may seem that any ZPD scheme might work. However, this is not true since it can become impractical for high-energy-consumption solutions due to the long delay, which is required to store sufficient energy.

For capacitor reservoirs, in order to determine the required capacitance, the energy available in the capacitor ($E_{cap}$), should be greater than the authentication energy ($E_{auth}$). The capacitance can be calculated using (\ref{eqn:e_available})~\cite{cypress2017}, where $V_{max}$ is the capacitor voltage when it is sufficiently charged and $V_{min}$ is when it has been used by the application or authentication process (see Figure~\ref{fig:imd-zpd-simple}).

\begin{equation}
\label{eqn:e_available}
E_{cap} = \frac{1}{2} \; C \; (V_{max}^2 - V_{min}^2) \; > \; E_{auth}
\end{equation}

RF-energy harvesters in general output \emph{constant power} instead of constant voltage~\cite{mishra2015charging}.
In this type of capacitor charging, the supplied voltage increases (instead of staying fixed) and current decreases with increasing capacitor voltage.
The capacitor charging time\footnote{The capacitor charging time for constant voltage charging is $5RC$.} ($t_{ch}$) for this type of charging is calculated using (\ref{eqn:charge_time})~\cite{mishra2015charging}. Here, $P_{ch}$ is the charging power supplied by the energy harvester to the capacitor ($C$), $R$ is the capacitor's equivalent series resistance (ESR) and $Q$ is the amount of coulombs stored during this time. Here $A = \sqrt{Q^2 + 4C^2RP_{ch}}$.

\begin{equation}
\label{eqn:charge_time}
%\begin{aligned}
t_{ch} = \frac{Q^2+QA+4C^2RP_{ch} \; \ln(\frac{A+Q}{\sqrt{4C^2RP_{ch}}})}{4CP_{ch}}
%\end{aligned}
\end{equation}

If the authentication-energy consumption is reduced then the required reservoir capacitance can be reduced as result.
If this value is within 0.1 $\mu F$ to 470 $\mu F$, then ceramic capacitors can be employed, which are ideal for energy harvesting because of low leakage current, small size and low cost~\cite{cypress2017}.
These capacitors also have a very low ESR~\cite{evanczuk2014capacitor}, which allows us to ignore the effect of the time constant ($RC$).
Hence, (\ref{eqn:charge_time}) can be simplified as (\ref{eqn:charge_time_reduced}), which is also equivalent to (\ref{eqn:charge_time_equivalent}). Here, $E$ is the energy stored in the capacitor.

\begin{equation}
\label{eqn:charge_time_reduced}
%\begin{aligned}
t_{ch} = \frac{Q^2}{2CP_{ch}}
%\end{aligned}
\end{equation}

\begin{equation}
\label{eqn:charge_time_equivalent}
%\begin{aligned}
t_{ch} = \frac{E}{P_{ch}}
%\end{aligned}
\end{equation}

The time it takes to charge an empty capacitor ($t_{{ch}_{initial}}$), and in the case of subsequent charging operations ($t_{{ch}_{repeat}}$) when a capacitor has a residue voltage of $V_{min}$ can be calculated by (\ref{eqn:charge_time_final})~\cite{cypress2017}. Here, $E_{initial} = \frac{1}{2} C V_{max}^2$, which is the energy attained by an empty capacitor when charged from 0 V to $V_{max}$.

\begin{equation}
\label{eqn:charge_time_final}
\begin{aligned}
t_{{ch}_{initial}} = \frac{E_{initial}}{P_{ch}} \\
t_{{ch}_{repeat}} = \frac{E_{cap}}{P_{ch}}
\end{aligned}
\end{equation}

\begin{figure}
	\centering
	\subfloat[Entering sleep based on voltage comparator interrupt]{\label{fig:vcmp-plot}{\includegraphics[trim={0.5cm 0.5cm 0.5cm 0.5cm},clip,scale=0.74]{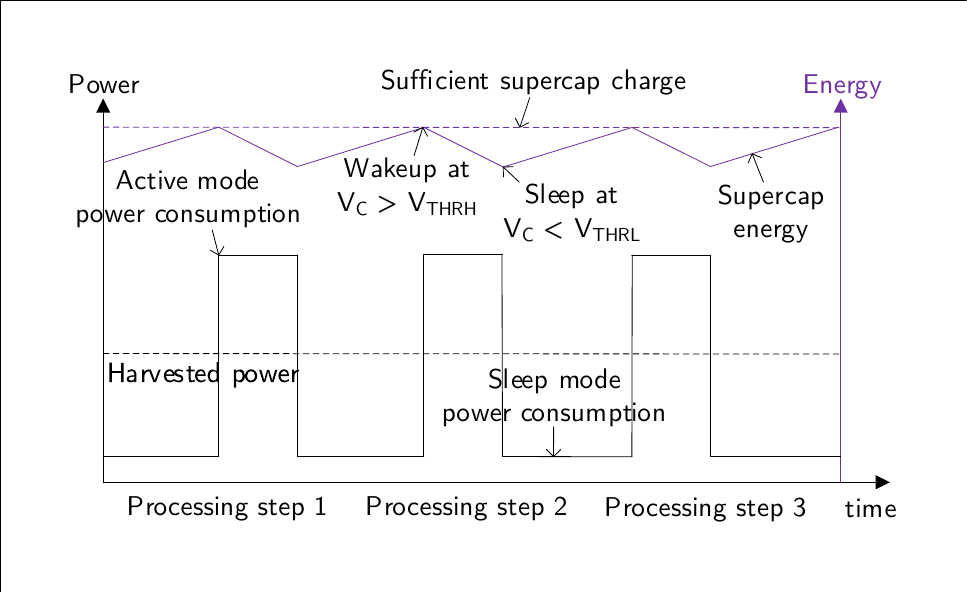} }}%
	\qquad
	\subfloat[Entering sleep after protocol step completion]{\label{fig:non-vcmp-plot}{\includegraphics[trim={0.6cm 0.5cm 0.6cm 0.5cm},clip,scale=0.74]{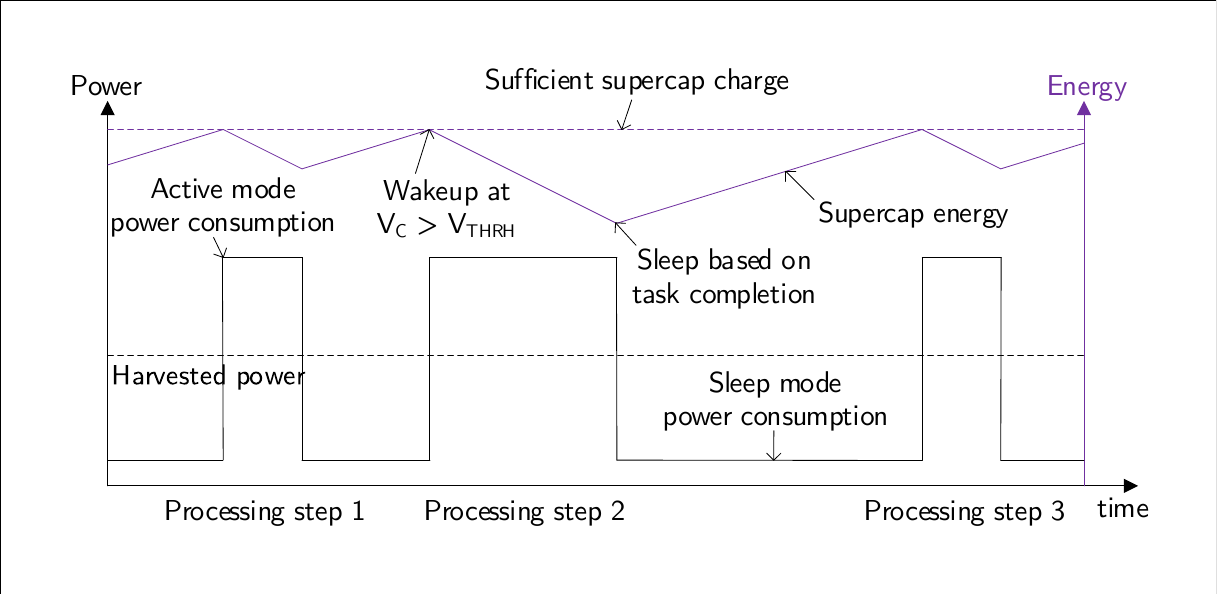} }}%
	\caption{Supercapacitor characteristics in relation to application duty cycle (active mode vs. sleep mode)}%
	\label{fig:duty-cycle-plot}%
\end{figure}

As an example, we use the evaluation setup from Section~\ref{sec:battery_size} and take the ISO/IEC 9798-2 based mutual authentication protocol from the ZPD solution in~\cite{strydis2013system}.
We use AES-128 for data confidentiality and cipher-based MAC.
For WPT, we look at the IPT scheme from~\cite{li2007wireless}, which is specifically designed for IMDs and delivers $P_{ch} = 6.15 \; mW$.
Using $V_{max} = 3.3 \; V$ and $V_{min} = 2.1 \; V$, which are within the operating supply voltage range of this setup (i.e., 2.05 $V$ to 3.5 $V$), we see that $C$ for the resulting scheme turns out to be 6.19 $\mu F$ (since the required $ E_{auth} = 20.07 \; \mu J$).
Using a standard ceramic capacitance of size greater than this value e.g., 10 $\mu F$, $t_{{ch}_{initial}}$ and $t_{{ch}_{repeat}}$ turn out to be 8.85 $ms$ and 5.27 $ms$ respectively, which are quite reasonable in terms of real-time behavior.

In general, the simplest solution is always to choose a reservoir capacitance that is much larger than the required value (as long as the charging delay is reasonable). This margin is important since the authentication protocol or the employed cryptographic primitives can change in the future, e.g., due to security updates.
However, in case $C$ turns out to be outside the ceramic-capacitor range due to large $E_{auth}$, we can employ the following schemes to reduce it, and the charging delay.

\subsubsection{Use of sleep modes}

The capacitor-charging delay can be minimized by using sleep modes and interrupts, instead of sizing the capacitor for the whole authentication, resulting in reduced required capacitance.
One way of achieving this could be to achieve a minimum required voltage ($V_{THR_H}$) using a voltage-controlled switch, before the capacitor energy is used by the rest of the IMD (Figure~\ref{fig:vcmp-plot}).
After some processing, the implant MCU can then enter sleep mode based on a voltage-comparator-based interrupt when the capacitor voltage ($V_C$) falls below a lower threshold ($V_{THR_L}$). Subsequently, the MCU can wakeup\footnote{These plots do not show the wakeup-time durations for clarity.} again if another such interrupt is set at $V_C > V_{THR_H}$~\cite{silabs2013}.
In this case, a \emph{protocol step}, such as a MAC calculation, can have multiple \emph{processing steps}.
Another way could be to go to sleep \emph{after} each protocol step in order to reduce the number of wakeups and the associated delay at the cost of a larger capacitor.
Here, the protocol step is the same as the processing step (Figure~\ref{fig:non-vcmp-plot}). In this case, the supercap size should be chosen based on the most energy-consuming protocol step.
However, this can be problematic if such a step is changed in the future due to the reprogramming of the IMD with a different authentication protocol.
Note that in this scheme as well the comparator interrupt will be required to wake up the device, indicating that the capacitor has been sufficiently charged.

\subsubsection{Gradual switch to harvested energy}

In another approach, the implant can use the battery for the first authentication request and if it fails, it can switch to harvested energy for subsequent accesses within a specified time-frame. This can allow for smaller reservoir sizes since we can afford the resulting delay due to frequent charge/discharge cycles in case of an inauthentic entity.

\subsection{Timeouts}
It can be argued that timeouts can be employed as a simpler alternative to ZPD. For instance, after a certain number of incorrect attempts, the IMD can be made to not accept further messages for a certain duration.
For domains other than IMDs this can be a natural choice. However, for IMDs, these timeouts can significantly compromise patient safety. For instance, any timeout after a malicious access can subsequently block a valid authentication attempt, which impacts \emph{availability}.

%% file: 900-conclusion.tex
\section{Conclusion}
\label{sec:conclusion}

Over the last few years, energy harvesting has been touted as a solution for protecting IMDs against battery-DoS.
In this paper, we have provided an extensive review of the IMD-specific ZPD works from literature.
We analyzed these works based on our formulated design considerations, and highlighted their shortcomings.
This paper is the first to substantiate these considerations and to provide specific recommendations towards practical ZPD implementations.
One strong recommendation is to employ adaptive ZPD in order to facilitate bedside-base-station operation.
As future work, we intend to develop a comprehensive ZPD scheme, which incorporates the lessons learned from this work.